\documentclass[5p,times,preprintnumbers,amsmath,amssymb,nofootinbib,superscriptaddress]{elsarticle}

\usepackage[T1]{fontenc}
\usepackage{lmodern}           
\usepackage{hyperref}

\usepackage{amsmath,amssymb,bm,mathrsfs}
\usepackage{upgreek}

\usepackage{graphicx}
\usepackage{caption}
\usepackage{subcaption}
\usepackage{tabularx}
\usepackage{dcolumn}
\usepackage{ragged2e}
\usepackage{float}
\usepackage{microtype}

\usepackage{xcolor}

\newcommand{\trento}{\texttt{ T$\mathrel{\protect\raisebox{-2.1pt}{R}}$ENT$\mathrel{\protect\raisebox{0pt}{o}}$}}

\begin{document}
	
	\title{Opacity estimation of OO collision from \texttt{CoMBolt-ITA hybrid}}%

	\author{S.~F.~Taghavi}
	\address{TUM School of Natural Sciences, Technische Universit\"at M\"unchen, Garching, Germany}
	
	\author{S.~M.~A.~Tabatabaee Mehr}
	\address{School of Particles and Accelerators, Institute for Research in
		Fundamental Sciences (IPM), P.O. Box 19395-5531, Tehran, Iran}

	\begin{abstract}

		Understanding the effect of system size on the applicability of the hydrodynamic description in heavy-ion physics remains unclear. Recent measurements of OO collisions at the LHC offer a new opportunity to refine our understanding of collectivity because of their intermediate size relative to heavy-ion and small-system collisions, as well as the relatively good control over their initial state. We use the \texttt{CoMBolt-ITA hybrid} model to describe recent OO measurements at the LHC. The model employs \trento{} for the initial state. A combination of the pre-equilibration and hydrodynamized medium stages is modeled consistently by \texttt{CoMBolt-ITA}, which evolves the Boltzmann distribution of massless collective excitations. The afterburner stage is included by employing \texttt{UrQMD}. Using this approach, we test whether the system lies in the regime where its spatial size approaches the mean free path, corresponding to low opacity, or in the opposite limit, where its size exceeds the mean free path sufficiently to enter the fluid-like evolution regime with high opacity. We find that, in light of the data-model comparison and considering the current status of the model, OO collisions with centralities larger than $60\%$ gradually leave the domain of fluid-like evolution.
		
	\end{abstract}
	
	\maketitle

	\textit{Introduction.---} For more than two decades, the observation of fluid-like signals, most notably elliptic flow, in heavy-ion collisions has been regarded as evidence for the formation of a collectively evolving QCD medium in the deconfined phase~\cite{STAR:2000ekf}. This observation has been connected to the flow-like behavior of the Quark--Gluon Plasma (QGP)~\cite{Ollitrault:1992bk}. Since then, many further measurements at RHIC and the LHC have been performed. These measurements have been described with remarkable accuracy by models in which hydrodynamics plays the central role in the collective evolution of the medium (see Ref.~\cite{Heinz:2024jwu} for a recent review).
	
	The main challenge emerged around 2010, when flow-like signals were observed in high-multiplicity pp collisions by the CMS Collaboration~\cite{CMS:2010ifv}. Similar observations were subsequently reported in other small systems (e.g., p--Pb, d--Au, etc.) at RHIC and the LHC~\cite{ALICE:2012eyl, ATLAS:2012cix, ALICE:2023lyr, STAR:2015kak, PHENIX:2018lia}. In recent years, comparable signatures have also been seen in e$^+$e$^-$ collisions~\cite{Belle:2022ars, Chen:2023njr}, $e^{-} p$ and $\gamma p$ collisions~\cite{ZEUS:2019jya, CMS:2022doq}, and even within jets~\cite{CMS:2023iam}.
	
	These observations naturally raise the question of how far a hydrodynamic description, as a long-wavelength effective theory, can be applied. In particular, it remains uncertain whether such an approach is still valid in systems that are both smaller in size and shorter in lifetime. To explore these limits, several microscopic or transport-based frameworks have been used as benchmarks or alternative descriptions (e.g., AMPT~\cite{Lin:2004en}). Models based on purely microscopic dynamics without explicit collectivity, such as PYTHIA~8 Angantyr~\cite{Bierlich:2018xfw}, also serve as useful references for estimating possible non-flow contributions. A natural way to further disentangle the underlying mechanisms is to study systems whose size is intermediate between that of large-ion collisions and the smallest collision systems. Such systems offer a regime where potential collective behavior can be probed while still testing the boundaries of hydrodynamics applicability.
	
	Recent measurements of OO and NeNe collisions at the LHC~\cite{ALICE:2025luc, ATLAS:2025nnt, CMS:2025tga} offer such an opportunity. Their system size falls between that of large-ion collisions (e.g., Pb--Pb) and small systems such as pp or p--Pb, providing an intermediate regime in which collectivity may be more cleanly probed. Moreover, unlike pp and p--Pb collisions, the initial-state geometry in OO and NeNe collisions is better constrained, thereby reducing uncertainties associated with limited knowledge of the initial state~\cite{Giacalone:2024luz}.
	
	The purpose of the present paper is to determine the nature of the medium produced in OO collisions based on the new LHC measurements, in particular those from the ALICE Collaboration. Specifically, we aim to assess whether the system size is comparable to the mean free path (particle-like regime) or considerably larger (fluid-like regime). Other studies have investigated the evolution of the medium by solving the Boltzmann equation in the relaxation-time approximation~\cite{Ferini:2008he,Ruggieri:2013ova,Romatschke:2015gic, Kurkela:2017xis, Kurkela:2019kip, Ambrus:2021fej, Ambrus:2022koq, Nugara:2024net}, including applications to OO collisions~\cite{Ambrus:2024hks}. In the current study, the parameters are tuned to describe the measured data, and by including the afterburner, we estimate the observables in kinematics that resemble the experimental measurements. After accounting for hadronization effects, we estimate the opacity parameter, which quantifies the particle-like or fluid-like properties of the system. In Ref.~\cite{Taghavi:2025xhl}, a new tool, \texttt{CoMBolt-ITA}, has been developed to solve the Boltzmann equation in 2+1 dimensions using a numerical scheme inspired by Ref.~\cite{Kurkela:2019kip}. Furthermore, a hybrid framework has been constructed by combining \trento{} initial conditions, freeze-out, and hadronic afterburner stages, enabling quantitative comparison of the simulations with experimental data. The code is publicly available in a GitHub repository~\cite{CoMBoltGithub}.
	
	In the following, we first review the conceptual background behind the implemented model. We then compare the model predictions with the data after tuning the relevant parameters. Afterwards, using a model constrained by data, we estimate whether the system is in a particle-like or fluid-like regime. Finally, we summarize our findings and outline our outlook.
	
	\textit{Quasiparticle evolution and opacity parameter.---}
	We now outline the main conceptual ingredients of \texttt{CoMBolt-ITA}. In this framework, the evolution of the system is described through the quasiparticle distribution $f(x,p)$ governed by the Boltzmann equation in the relaxation-time approximation,
	\begin{equation}
		p^\mu \partial_\mu f(x,p) = -\frac{f - f_{\text{eq}}}{\tau_{\text{relax}}}.
	\end{equation}
	Starting from $\tau_0 = 0^+$, the early evolution of the system is approximately free-streaming until it hydrodynamizes at $\tau_* \sim \tau_{\text{relax}}$. The longitudinal expansion during this stage leads to the relation $\epsilon_0 \tau_0 \sim \epsilon_* \tau_*$, where $\epsilon_*$ denotes the energy density at the onset of hydrodynamization. At sufficiently high initial energies, the system approximately exhibits conformal symmetry, implying $\tau_{\text{relax}} = \gamma^{-1} (\epsilon_*)^{-1/4}$, where $\gamma$ is a dimensionless parameter. Using these inputs, one obtains $\tau_{\text{relax}} = \gamma^{-4/3} (\epsilon_0 \tau_0)^{-1/3}$. 
	The system size $R$ measured in units of the mean free path $\tau_{\text{relax}}$ can then be expressed as $R / \tau_{\text{relax}} = \hat{\gamma}^{4/3}$, where $\hat{\gamma} = \gamma\, R^{3/4} (\epsilon_0 \tau_0)^{1/4}$ is the opacity parameter~\cite{Kurkela:2019kip}. In practice, $\epsilon_0$ is replaced by the average of the total energy density per unit pseudorapidity in the transverse plane, $\epsilon_0 \sim \epsilon_{\text{tot}} / (\pi R^2)$, leading to the definition used in this work,
	\begin{equation}
		\hat{\gamma} = \gamma\, ( R\, \epsilon_{\text{tot}} \tau_0 / \pi )^{1/4},
	\end{equation}
	where $\epsilon_{\text{tot}} = \{1\}$ and $R = \{x^2 + y^2\} / \epsilon_{\text{tot}}$. We use the notation
	\begin{equation}
		\{ g(x_\perp) \} \equiv \int dx_\perp\, g(x_\perp)\, \epsilon_0(x_\perp),
	\end{equation}
	with $\epsilon_0(x_\perp)$ the initial transverse energy density distribution.
	
	When $\tau_{\text{relax}} \gtrsim R$ ($\hat{\gamma} \lesssim 1$--$2$), quasiparticles undergo only a few interactions and the system remains in the \textit{particle-like} regime. In contrast, for $\tau_{\text{relax}} < R$ ($\hat{\gamma} \gtrsim 4$), the system enters the \textit{fluid-like} regime~\cite{Kurkela:2019kip}. The specific numerical thresholds for $\hat{\gamma}$ follow the convention used to characterize the point at which the development of momentum anisotropy from a given spatial anisotropy begins to deviate substantially from that obtained in the ideal hydrodynamic limit. In particular, it has been demonstrated in Ref.~\cite{Kurkela:2019kip} that the difference between the energy--momentum tensor obtained from kinetic theory and that constructed from hydrodynamic equations with second-order corrections in the derivative expansion becomes considerably large in the particle-like regime, while the difference remains small in the fluid-like regime.  In the particle-like regime, the system has a very limited ability to generate collective flow.

	\textit{Collective model \texttt{CoMBolt-ITA hybrid}.---} We now test the above concepts in OO collisions. For this purpose, a realistic initial state is required, together with a treatment of hadronization and the subsequent hadronic cascade. In the following, we briefly describe the model framework covering all stages of the collision.
	
	We assume that the evolution starts at $\tau_0 = 0.1~\text{fm}/c$. Since the massless quasiparticles are assumed, only the magnitude of the quaiparticles' momentum is fixed and can be integrated out which leads to an integrated version of $f(x,p)$ shown by $F(\tau,x_\perp, v_z, \phi_p)$ at which $(v_z, \phi_p)$ contains the angular part of the quasiparticles' momentum. It means
	the initial Boltzmann distribution requires both spatial and momentum information, which in this study are taken to factorize, as $F(\tau_0,x_\perp, v_z, \phi_p) = 2\epsilon_0(x_\perp) \mathcal{P}_0(v_z, \phi_p)$. For the spatial profile, $\epsilon_0(x_\perp)$, the initial state is generated using the \trento{} model~\cite{Moreland:2014oya}. For the nucleon configurations, we employ those provided in Ref.~\cite{Giacalone:2024luz}. 
	The nucleon--nucleon inelastic cross section at the given center-of-mass energy determines which nucleons participate in the collision. The participating nucleons then generate the thickness functions $T_{A}$ and $T_{B}$ for the projectile and target, respectively. Each nucleon is modeled with a Gaussian profile of width $w = 0.6\;$[fm/$c$] and is assumed to consist of $n_c = 4$ subnucleonic constituents, each of width $v = 0.46$. The contribution of each constituent to the local thickness is randomly rescaled by a factor sampled from a Gamma distribution with unit mean value and shape parameter $k = 1.92$. Once the participant thickness functions are constructed, they are combined phenomenologically to form the reduced thickness function $T_R$, which we use as the initial energy density profile. The parameter $p$ in the \trento{} model is taken to be $p = 0.0$. These parameterizations are   consistent with Bayesian analyses based on relativistic hydrodynamic simulations~\cite{Bernhard2019,JETSCAPE:2020mzn,Nijs:2020roc,Parkkila:2021tqq,Parkkila:2021yha}, but are not chosen particularly from an specific study.
	
	For the momentum distribution, we assume a smooth form,
	\begin{align}\label{momentumDist}
		\mathcal{P}_0(v_z, \phi_p)
		= \bigg( 2 \lambda\, \tan^{-1}[\sinh(1/\lambda)] \, \cosh\!\left(\frac{v_z}{\lambda}\right) \bigg)^{-1},
	\end{align}
	characterized by a parameter $\lambda$. In the limit $\lambda \to 0$, the longitudinal pressure vanishes, while for $\lambda \to \infty$ the initial pressure becomes isotropic. Thus, $\lambda$ is a model parameter to be constrained by data. In this work, we fix $\lambda = 0.1$.

	In the approach presented in this work, there is no need to include a separate pre-equilibrium stage, since \texttt{CoMBolt-ITA} is applicable in highly anisotropic regimes. The energy–momentum tensor is obtained as the second moment of the Boltzmann distribution in momentum space, and the energy density and collective velocity are then determined through Landau matching. For massless quasiparticles, the system corresponds to a medium with an ideal equation of state, $\epsilon = C_0 T^4$, and vanishing bulk viscosity. In our calculation, the collision kernel is proportional to $\gamma = \tau_{\text{relax}}^{-1},\epsilon^{-1/4}$, where $\tau_{\text{relax}} = 5(\eta/s)/T_{\text{eff}}$ and $\eta/s$ is the shear viscosity to entropy density ratio. To partially account for the QCD equation of state, we have added an option to modify the collision kernel by replacing $C_0$ with $C_0(T_{\text{eff}})$, as reported by the HotQCD Collaboration~\cite{HotQCD:2014kol}. Using a temperature-dependent $C_0$ introduces a mild dependence of $\gamma$ and consequently $\hat{\gamma}$ on the transverse position. This implies that the collision kernel in hotter regions will have a larger $C_\epsilon$ compared to regions close to the switching description from quasiparticles to actual hadrons. Ultimately, we found that a reasonable description of the data is obtained by using $C_\epsilon = 4.27$, which is close to the value of $C_0(T_{\text{eff}})$ in the switching region.

	The description switching (freeze-out) surface is constructed using the Cornelius algorithm~\cite{Huovinen:2012is}, with a switching temperature of $T_{\text{sw}} = 0.155~\text{GeV}$. For particlization, the energy-momentum tensor at each time step is evaluated and passed to the freeze-out code developed by the Duke group~\cite{Bernhard:2018hnz}. The code constructs the Boltzmann distribution of the hadron gas according to the prescription developed by Pratt and Torrieri~\cite{Pratt:2010jt}. The hadronic gas evolves through the \texttt{UrQMD} afterburner~\cite{Bass:1998ca,Bleicher:1999xi}.
	
	For the overall initial-state normalization and $\eta/s$, we present two scenarios (see Table~\ref{tab:parameters}): one with a large $\eta/s = 0.18$ and a large normalization of $70$, and another with a small $\eta/s = 0.1$ and a small normalization of $50$. These choices are made to reproduce the elliptic flow $v_2\{2\}$ as a function of centrality. For larger $\eta/s$, more evolution time is needed to generate sufficiently large anisotropic flow, which requires a higher initial total energy density achieved through the higher normalization. The normalization in this study is higher compared to that obtained from Bayesian analyses. This difference arises because particlization in those studies begins after the pre-equilibrium stage (see e.g. Ref.~\cite{Nijs:2020roc}), whereas here it starts earlier, within the pre-equilibrium stage.
	
	\begin{table}[t!]
		\centering
		\begin{tabular}{l c c}
			\hline
			& \hspace{1cm} Norm \hspace{1cm} & \hspace{1cm} $\eta/s$  \hspace{1cm} \\
			\hline
			Case 1 & \;50 & 0.1 \\
			Case 2 & \;70 & 0.18 \\
			\hline
		\end{tabular}
		\caption{Two selected parameterizations.}\label{tab:parameters}
	\end{table}

	\textit{Event-by-event simulation.---} We have around $3\times 10^3$ minimum bias events each oversampled $10^3$ times. The centrality classes are determined based on the charged-particle multiplicity in unit pseudorapidity. 
	
\begin{figure}[t!]
		\centering
		
		\begin{subfigure}{1.0\linewidth}
			\centering
			\includegraphics[width=\linewidth]{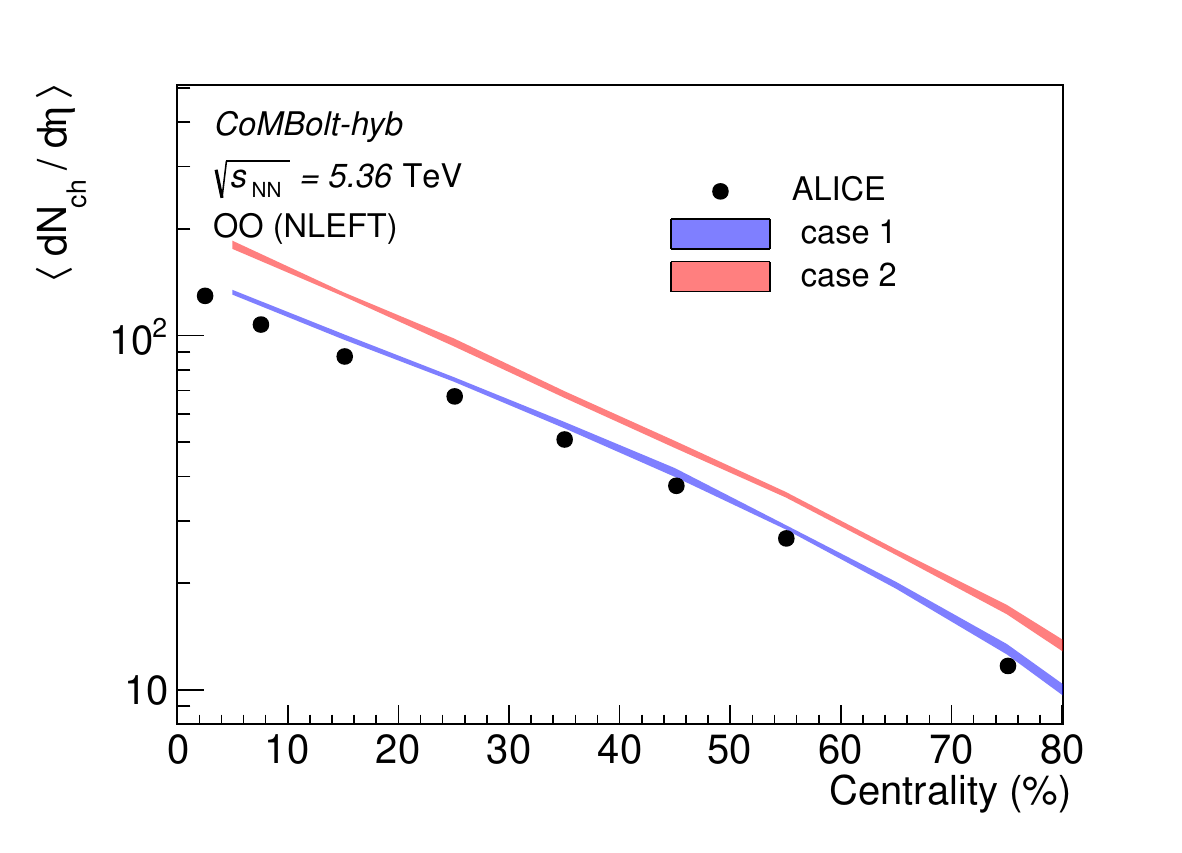}
		\end{subfigure}
		
		\vspace{0.0cm}
		
		\begin{subfigure}{1.0\linewidth}
			\centering
			\includegraphics[width=\linewidth]{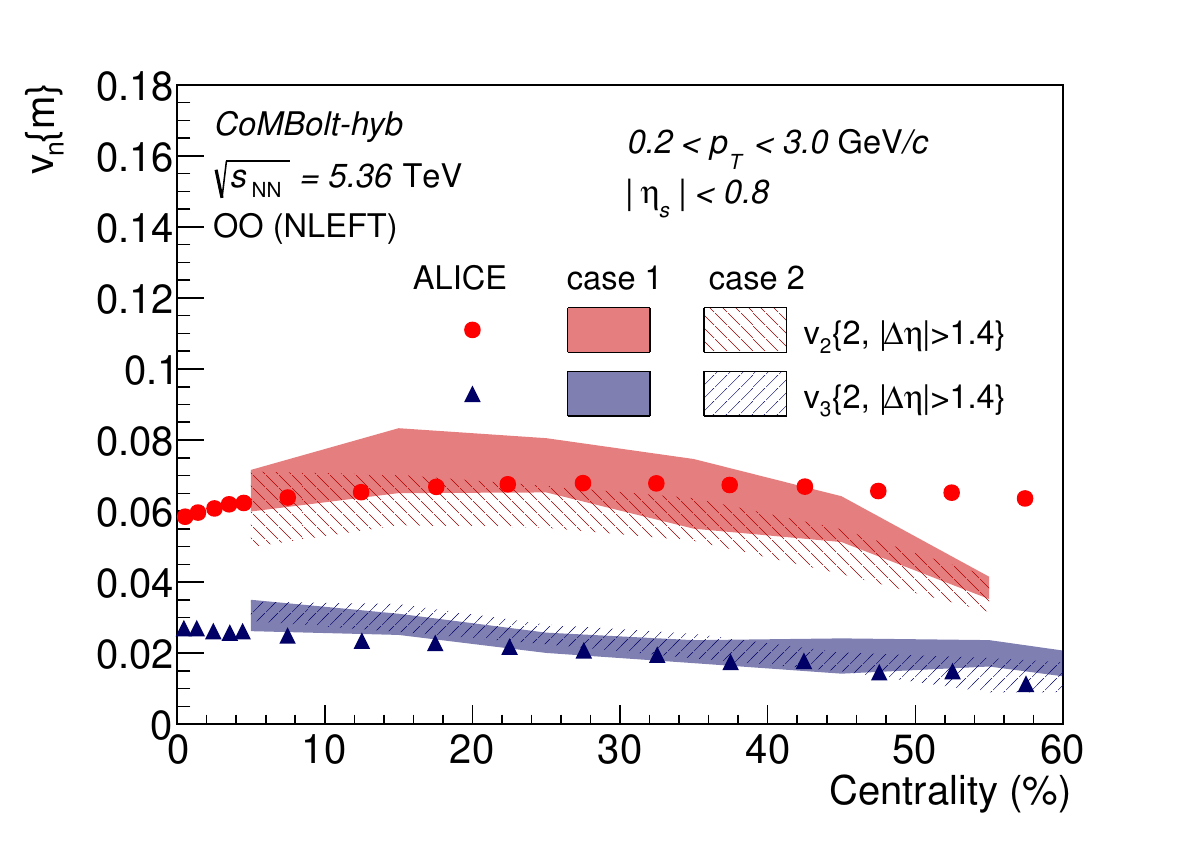}
		\end{subfigure}
		
		\caption{\justifying
			Charged particle multiplicity in unit pseudorapidity from ALICE preliminary (presented at the Initial Stages Conference 2025) compared to \texttt{CoMBolt-ITA} hybrid (top).  
			Two-particle correlations of elliptic and triangular flow from the ALICE Collaboration~\cite{ALICE:2025luc} (filled points) and \texttt{CoMBolt-ITA} hybrid as a filled band (bottom).
		}
		\label{fig:model_data}
	\end{figure}

	In Fig.~\ref{fig:model_data}, we compared the \texttt{ComBolt} outcome with two different paranmeterizations with ALICE measurement. The charged multiplicity in the unit rapidity, $dN_{\text{ch}}/d\eta$, corresponding to each centrality class is shown in Fig.~\ref{fig:model_data} (top). As discussed before, since the larger $\eta/s$ (case 1) requires more time to develop $v_2$, a larger normalization is needed to increase the initial energy density and consequently leads to a larger multiplicity.

	For the elliptic and triangular flow coefficients, we use the standard multiparticle correlation method~\cite{Bilandzic:2010jr,Bilandzic:2013kga} with an $\eta$-gap~\cite{Jia:2017hbm}. The pseudorapidity range $\eta_s$ is divided into two subevents, $A$ and $B$, with $\eta_s \in (-1.6,\,-1.4)$ for $A$ and $\eta_s \in (1.4,\,1.6)$ for $B$. The transverse momentum of the selected particles is restricted to $0.2 < p_T < 3.0~\text{GeV}$. These choices follow those used in the ALICE measurements.
	
	The $Q$-vectors for charged hadrons are defined as
	\begin{equation}
		Q_{n,A(B)} = \sum_{j \in A(B)} e^{i n \varphi_j},
	\end{equation}
	where $\varphi_j$ is the azimuthal angle of particle $j$. The two-particle cumulant flow coefficients are then obtained as
	\begin{equation}
		v_n\{2\} = \sqrt{\frac{Q_{n,A} Q_{n,B}^*}{M_A M_B}},
	\end{equation}
	where $M_{A(B)}$ is the multiplicity in subevent $A(B)$.
	
	The results are shown in Fig.~\ref{fig:model_data} (bottom) for two different cases, which are rather compatible, as we demanded from the beginning. As seen, the model reproduces the data reasonably well up to the $40$--$50\%$ centrality region. We emphasize two points. First, before particlization, the model evolves the system using massless quasiparticles, corresponding to an ideal EoS. Including bulk viscosity and a realistic QCD EoS may modify the results. Second, the model parameters require tuning. As indicated by Bayesian analyses of relativistic hydrodynamics, variations in these parameters can lead to changes in the final observables.
	
	As seen, the case 1 parameterization is more compatible with the data. Considering that the parameter tuning yields results approximately compatible with the experimental data, we proceed to evaluate the collectivity response of the system as a function of the opacity parameter. Based on the tuned parameters, we can assess whether the medium produced in OO collisions behaves in a particle-like or fluid-like manner.

	\begin{figure}
		\includegraphics[width=1.0\linewidth]{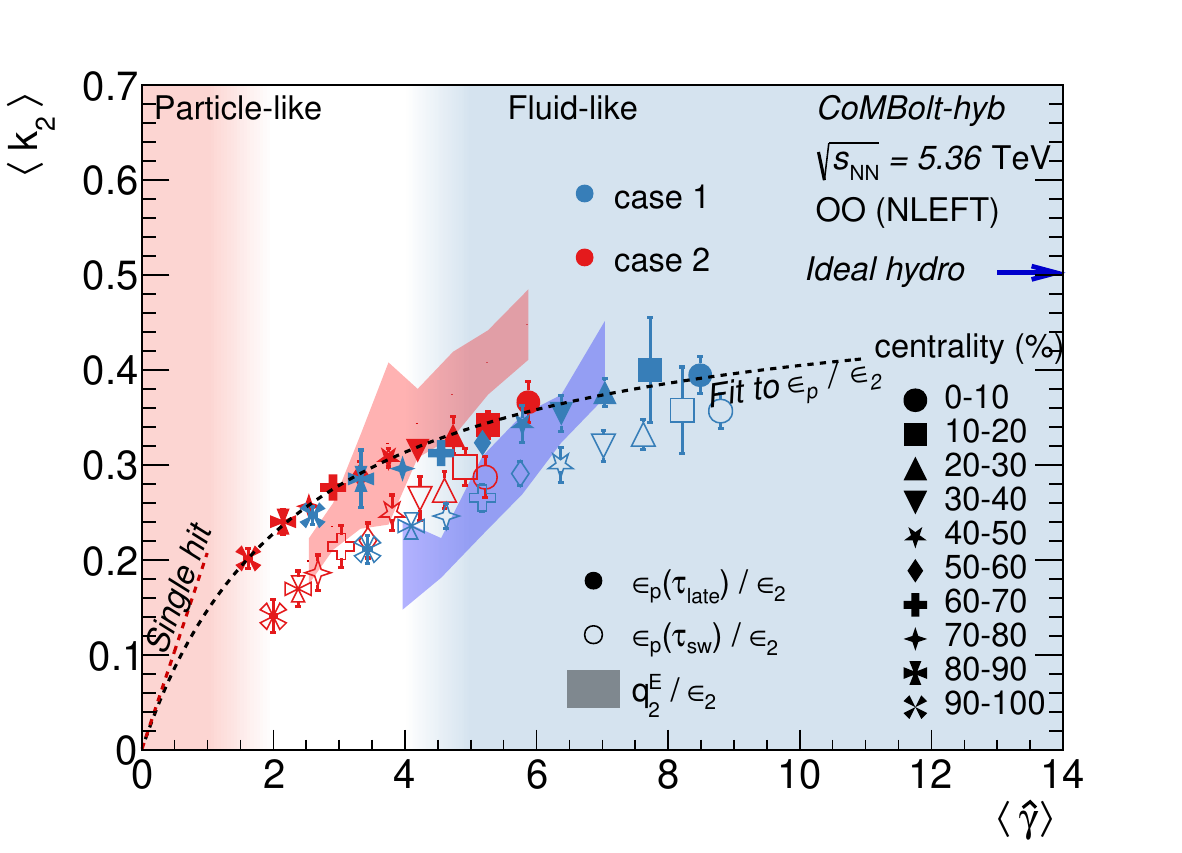}
		\caption{\justifying Collectivity response as a function of opacity for various centralities using three different measures: momentum anisotropy at late times (filled markers), momentum anisotropy at the switching time (open markers), and transverse-momentum elliptic flow after the hadronic cascade (filled band). Case 1, which shows better agreement with data, indicates that OO collisions at centralities below $60\%$ are more consistent with fluid-like evolution.}
		\label{fig:k2}
	\end{figure}
	
	\textit{Discussion.---} To estimate the magnitude of the collectivity response, we use three different measures. The anisotropic flow is non-vanishing if the initial spatial anisotropy is translated to the momentum anisotropy. The initial spatial anisotropy at each event can be quantified by 
	\begin{equation}
		\epsilon_2 =\left| -\{ r^2 e^{2i\Phi} \} / \{r^2\}\right|,
	\end{equation}
	where $(r,\Phi)$ is the spatial coordinate in the transverse direction in polar coordinates. A measure for the momentum anisotropy is obtained from the energy--momentum tensor,
	\begin{equation}\label{eq:localMomentumEllip}
		\begin{split}
			&\epsilon_p(\tau) = \\
			&\left| \frac{\displaystyle \int dx_\perp \left[ T^{xx}(\tau, x_\perp) - T^{yy}(\tau, x_\perp) + 2i\, T^{xy}(\tau, x_\perp) \right]}
			{\displaystyle \int dx_\perp \left[ T^{xx} + T^{yy} \right]} \right|.
		\end{split}
	\end{equation}
	
	The first measure of quantifying the collectivity response is the ratio $k_2 = \epsilon_p(\tau_\text{late}) / \epsilon_2$, where $\tau_\text{late}$ is chosen such that $k_2$ for central OO collisions approaches a constant value. In this case, the response coefficient is calculated from a scale-free system, and a scaling behavior is expected. For each centrality class, the averages of $\hat{\gamma}$ and $k_2$ are evaluated, and the results are shown in Fig.~\ref{fig:k2} (filled symbols). We observe that the two different parameterization follow the same trend, approaching zero in the small-opacity limit. 
	
	We now discuss whether switching the description affects this scaling behavior. It has been suggested in Ref.~\cite{Kurkela:2019kip} to examine the transverse-energy elliptic flow, which is expected to be less modified during hadronization,
	\begin{equation}
		\frac{dE_T}{dp_T\, d\eta\, d\phi}
		= \frac{1}{2\pi} \frac{dE_T}{dp_T\, d\eta}
		\left( 1 + 2\sum_{n=1}^\infty v_n^E \cos\left[n(\phi - \psi_n^E)\right] \right),
	\end{equation}
	where $E_{T,i} = E_i\, p_{T,i} / |\vec{p}_i|$ is the transverse energy of particle~$i$. From this expression, one can estimate the transverse anisotropic flow as
	\begin{equation}\label{energyFlow}
		v_n^E e^{in\psi_n^E}
		\approx \hat{q}^E_n
		\equiv
		\frac{\displaystyle \sum_i E_{T,i}\, e^{in\varphi_i}}
		{\displaystyle \sum_i E_{T,i}},
	\end{equation}
	where the sum runs over charged hadrons. This quantity should approximately represent the full hadron spectrum (charged + neutral), since the numerator and denominator are expected to scale similarly.
	
	The reduced $q$-vector weighted by transverse energy, Eq.~\eqref{energyFlow}, provides the second measure of the response coefficient and includes the effects of the hadronic afterburner. The corresponding values of $k_2$, extracted from $q_2^E$, are shown as shaded bands in Fig.~\ref{fig:k2}. As seen, the response coefficient no longer exhibits a scaling behavior. This breakdown is expected, since hadronization and hadronic rescattering explicitly violate the scale-free evolution assumed in the pre-hadronic stage.
	
	It is important to note that this violation of scaling does not imply that systems of different sizes (pp, pPb, OO, NeNe, PbPb) would exhibit such a large scaling violation, because, in principle, a single set of parameters should describe the data. What we observe here is that in a system with a characteristic scale, the strength of the collectivity depends on the initial total energy density and $\eta/s$, which leads to the parameters used in case 1 and case 2 not following the same trends.
	
	We also examined whether the switching time has a significant effect on the scaling violation. To address this, we used a third measure to estimate $k_2$. Specifically, we compute $\epsilon_p(\tau_{\text{sw}})$, where $\tau_{\text{sw}}$ is not a fixed late time but the moment when the last medium cell is switched to hadron gas. The corresponding values of $k_2$ are shown as open symbols in Fig.~\ref{fig:k2}. Consequently, for smaller systems and more peripheral collisions, $\epsilon_p$ is evaluated at earlier times, resulting in less development of momentum anisotropy. In this definition, $\epsilon_p$ explicitly depends on a characteristic timescale, $\tau_{\text{sw}}$. 
	
	We find that $k_2$ is systematically lower in both cases of parameterization compared to $\epsilon_p$ calculated at late times. This indicates that the anisotropy generated during the hadronic stage is not the same as that produced in the deconfined phase. Although a signature of scaling violation appears here, the overall scaling behavior remains, suggesting that another mechanism must be responsible for the scaling violation in $k_2 = q_2^E/\epsilon_2$.
	
	\begin{figure}[t!]
		\includegraphics[width=1.0\linewidth]{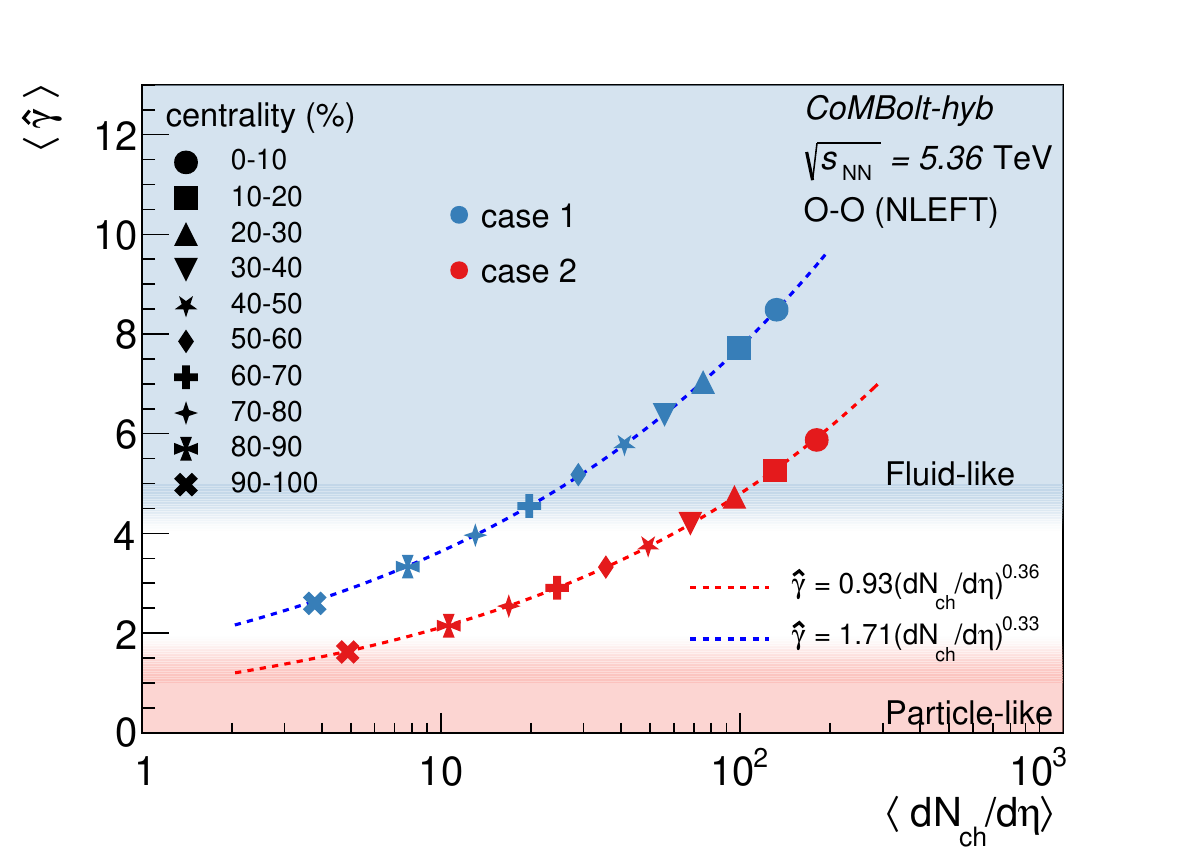}
		\caption{\justifying Opacity as a function of charged particle multiplicity. Case 1, which was previously shown to have better agreement with ALICE measurements, indicates that OO collisions at centralities below $60\%$ are more consistent with fluid-like evolution.}
		\label{fig:dnch}
	\end{figure}
	
	We finally present the average opacity as a function of the average charged multiplicity per unit pseudorapidity for different centralities in Fig.~\ref{fig:dnch}. From this plot, one observes a clear connection between the opacity parameter and the charged multiplicity. One can estimate the behavior of the opacity parameter $\hat{\gamma}$. In Ref.~\cite{Giacalone:2019ldn}, the final multiplicity is related to the initial-state energy via $dN_\text{ch}/d\eta \propto (\eta/s)^{1/3} (\epsilon_0 \tau_0)^{2/3}$. Combining this with the definition of the opacity parameter, one obtains\footnote{This estimate is compatible with that of Ref.~\cite{Arslandok:2023utm}, where a different convention for the opacity is used, $\chi = \hat{\gamma}^{8/9}$.}
	\begin{equation}
		\hat{\gamma} \propto \frac{1}{(\eta/s)^{1.125}}
		\left( \frac{dN_\text{ch}}{d\eta} \right)^{0.375}.
	\end{equation}
	As seen in Fig.~\ref{fig:dnch}, there is good agreement between the $dN_\text{ch}/d\eta$ power extracted from the fit and that estimated from Ref.~\cite{Giacalone:2019ldn}.
	
	The anisotropic flow and charged multiplicity from \texttt{CoMBolt} and conventional hydrodynamic codes show quite similar agreement between model and data~\cite{ALICE:2025luc, ATLAS:2025nnt, CMS:2025tga}. However, one should also consider other observables such as the $p_T$ spectrum and the average $p_T$. Although these observables have not yet been reported, we have compared our results with hydrodynamic simulations. With the current parameter tuning, we observe systematically larger average $p_T$. By examining the hydrodynamic code \texttt{VISH2+1}, we find that a major contribution to this difference arises from the combination of a non-ideal EoS and bulk viscosity, which reduces the average $p_T$ by roughly 30\%, in agreement with the estimation in Ref.~\cite{Gardim:2019xjs}. This highlights the necessity of including the mass scale in the system.

	\textit{Conclusion.---} Based on the current status of the model, the recent LHC measurements in OO collisions show reasonably good agreement with \texttt{CoMBolt} for the scenario with small $\eta/s$ (case 1). With access to initial-state information from theory, and by examining the collectivity response $k_2$ as a function of opacity, as well as opacity as a function of charged particle multiplicity, we observe that OO collisions with centrality below $60\%$ lie within the regime of fluid-like evolution, although the actual numerical values remain affected by the idealizations of the code and the absence of a systematic parameter determination through Bayesian tuning.
	
	The case 2 parameterization can also reproduce the $v_2\{2\}$ data by construction, suggesting a more particle-like dynamical regime. However, as discussed, this parameterization fails to describe the charged particle multiplicity. This indicates that anisotropic flow alone is insufficient for a robust determination of the dynamical regime. Additional experimental observables, in our case, the charged particle multiplicity, must be included in the parameter tuning to obtain a more reliable estimation. This reflects the fact that different dynamical scenarios can lead to similar flow observables, making additional constraints necessary.

	\textit{Summary and outlook.---} We have used the \texttt{CoMBolt-ITA hybrid}  to study the recently measured OO collisions at the LHC, in particular those reported by the ALICE experiment. We found that the data favor a system with a small $\eta/s = 0.1$. With the parameters anchored to data, we examined whether the system is more compatible with a fluid-like or particle-like dynamical regime. Based on the current status of the model, we find that OO collisions at centralities smaller than $60\%$ lie within the range of fluid-like dynamics.
	
	A substantial development is required to include a mass term for the quasiparticles, corresponding to a non-ideal equation of state with non-vanishing bulk viscosity. This requires solving the full Boltzmann equation rather than the momentum-magnitude--integrated version, as has been discussed so far. The former also enables the inclusion of a collision kernel based on QCD. Recently, a method based on machine learning has been introduced in Ref.~\cite{BarreraCabodevila:2025ogv} to evaluate the effective kinetic theory collision kernel. Examining other collision systems, such as pp collisions, is also planned. One can also ask whether multi-particle correlations associated with higher orders of the BBGKY hierarchy can have a measurable impact on final-state observables. Initial attempts to go beyond the one-particle distribution function have been made by incorporating multi-particle correlations through extensions of the BBGKY hierarchy, for instance within the correlation time approximation (CTA) framework~\cite{Grozdanov:2024fxr}. 
	

	\textit{Acknowledgment.---} We would like to thank Farid Taghinavaz for valuable discussions. S.F.T. also thanks Aleksi Kurkela and Giuliano Giacalone for insightful discussions. Part of this work was carried out while S.F.T. attended the 2025 MIAPbP workshop “Event Generators” at the Munich Institute for Astro-, Particle and BioPhysics (MIAPbP), which is funded by the Deutsche Forschungsgemeinschaft (DFG) under Germany’s Excellence Strategy – EXC-2094 – 390783311. The final stages of this work were completed during the workshop “Light-Ion Collisions at the LHC 2025” at CERN, whose stimulating environment is gratefully acknowledged. S.F.T. is supported by the DFG through grant number 517518417.

	\bibliographystyle{unsrt}  
	\bibliography{references.bib}%
	
\end{document}